Original Paper

# An Urban Population Health Observatory System to Support COVID-19 Pandemic Preparedness, Response, and Management: Design and Development Study

Whitney S Brakefield[1,2], BSc, MSc; Nariman Ammar[2], PhD; Olufunto A Olusanya[2], PhD; Arash Shaban-Nejad[2], MSc, PhD, MPH

[1]Bredesen Center for Data Science and Engineering, University of Tennessee, Knoxville, TN, United States
[2]Center for Biomedical Informatics, Department of Pediatrics, College of Medicine, University of Tennessee Health Science Center, Memphis, TN, United States

**Corresponding Author:**
Arash Shaban-Nejad, MSc, PhD, MPH
Center for Biomedical Informatics, Department of Pediatrics, College of Medicine
University of Tennessee Health Science Center
50 N Dunlap Street
Memphis, TN, 38103
United States
Phone: 1 901 287 583
Email: ashabann@uthsc.edu

## Abstract

**Background:** COVID-19 is impacting people worldwide and is currently a leading cause of death in many countries. Underlying factors, including Social Determinants of Health (SDoH), could contribute to these statistics. Our prior work has explored associations between SDoH and several adverse health outcomes (eg, asthma and obesity). Our findings reinforce the emerging consensus that SDoH factors should be considered when implementing intelligent public health surveillance solutions to inform public health policies and interventions.

**Objective:** This study sought to redefine the Healthy People 2030's SDoH taxonomy to accommodate the COVID-19 pandemic. Furthermore, we aim to provide a blueprint and implement a prototype for the Urban Population Health Observatory (UPHO), a web-based platform that integrates classified group-level SDoH indicators to individual- and aggregate-level population health data.

**Methods:** The process of building the UPHO involves collecting and integrating data from several sources, classifying the collected data into drivers and outcomes, incorporating data science techniques for calculating measurable indicators from the raw variables, and studying the extent to which interventions are identified or developed to mitigate drivers that lead to the undesired outcomes.

**Results:** We generated and classified the indicators of social determinants of health, which are linked to COVID-19. To display the functionalities of the UPHO platform, we presented a prototype design to demonstrate its features. We provided a use case scenario for 4 different users.

**Conclusions:** UPHO serves as an apparatus for implementing effective interventions and can be adopted as a global platform for chronic and infectious diseases. The UPHO surveillance platform provides a novel approach and novel insights into immediate and long-term health policy responses to the COVID-19 pandemic and other future public health crises. The UPHO assists public health organizations and policymakers in their efforts in reducing health disparities, achieving health equity, and improving urban population health.

*(JMIR Public Health Surveill 2021;7(6):e28269)*   doi: 10.2196/28269

## Introduction

### Background

COVID-19 is a highly transmissible disease caused by SARS-CoV-2. COVID-19 has been one of the leading causes of death in many countries since December 2019 when it was first reported in Wuhan, China. As of May 5, 2021, there have been over 156 million confirmed cases of COVID-19 and more than 3.2 million deaths worldwide [1,2]. Currently, the United States is among the countries leading in the number of confirmed COVID-19 cases and deaths. According to the Institute for Health Metrics and Evaluation, by August 2021, there will be an estimated 600,000 COVID-19–related deaths in the United States [3]. COVID-19 illness can range from asymptomatic and mild to moderate and severe, with symptoms that include cough, shortness of breath, sore throat, fever, fatigue, and muscle pain. Underlying comorbid health conditions such as hypertension, diabetes, chronic obstructive pulmonary disease, cardiovascular disease, and cerebrovascular disease increase the risk of severe complications from COVID-19. These complications include acute respiratory failure, pneumonia, acute kidney or liver injury, blood clots, and possibly death [4,5].

Although COVID-19 adversely affects people's lives in many respects, it disproportionally impacts certain groups and populations [6-11]. Researchers [12] are still exploring the dynamics of the COVID-19 pandemic in urban areas to understand the short- and long-term impacts of COVID-19 on urban environments, which may be densely populated. The socioeconomic and environmental risk factors and determinants are key to explain how urban life affects population health [13]. Underlying sociocontextual factors such as Social Determinants of Health (SDoH) could increase the prevalence of COVID-19 and COVID-19–related deaths in certain communities. A relatively new concept in health care, SDoH are defined by the World Health Organization as "the conditions of where a person is born, where they grow up, where they live, where they work, and where they age" [14]. SDoH is comprised of five domains: economic stability, education access and quality, neighborhood and built environment, health care access and quality, and social and community context [15,16]. Figure 1 illustrates several SDoH variables that are included in each domain, which are as follows. The economic stability domain encompasses the level to which an individual or group falls within the hierarchical societal structure, with variables that reflect the impact of socioeconomic conditions. The education access and quality domain consists of variables that influence the process of becoming educated. The neighborhood and built environment domain consists of variables that relate to the physical environment and have the potential to overlap with other domains, making them some of the most flexible domains. The health care access and quality domain include variables that reflect access to health care and describe how health information is interpreted to make appropriate and informed decisions. The social and community context domain encompasses variables that demonstrate the social setting an individual resides in and their community involvement. These SDoH are impacted by access to power, money, and other resources and are therefore considered to be the major driving force behind health inequities [14].

**Figure 1.** Five domains and variables of the Social Determinants of Health.

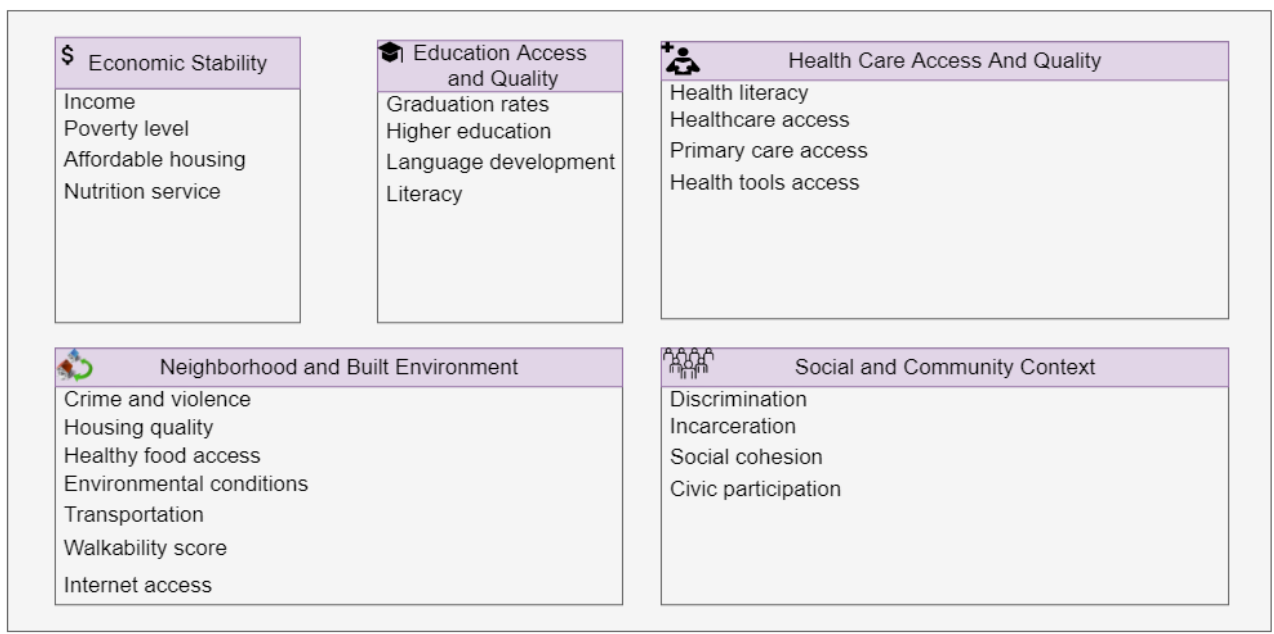

Our prior studies have explored associations between SDoH and several health outcomes [17-19]. Consequently, our findings underpin the importance of incorporating SDoH when implementing intelligent public health surveillance solutions to inform public health decisions. During the COVID-19 pandemic, the development of an intelligent surveillance platform that embeds SDoH indicators can improve equity in the uneven distribution of quality health care services (ie, testing and vaccination), inform health officials on the timing, phasing, and safety for reopenings, and address shortages of medical supplies, devices, and health care workers to alleviate the related health and economic crisis [20,21]. Such a platform would serve as an apparatus for actualizing effective intervention design and implementation to address health disparities, provide awareness to the general public, and improve public health decision-making and planning. Additionally, it can foster the consistent integration of surveillance data across jurisdictions to estimate the incidence and prevalence of different health conditions and related risk factors. Finally, it can be used for intelligent query-answering to formally interrogate hypothesis-driven research questions.

### Objectives

In this study, we describe the design and development of the Urban Public Health Observatory (UPHO), a web-based knowledge-based surveillance platform that integrates multidimensional heterogenous data including SDoH indicators and population health data and provides near–real-time analysis and dashboarding of ongoing COVID-19–related comorbidities and mortalities. In particular, we aimed to (1) redefine the Healthy People 2030's SDoH taxonomy to classify SDoH indicators into 6 domains to characterize barriers that are specific to the COVID-19 pandemic and (2) design and develop a prototype for the UPHO surveillance platform.

## Methods and Results

### Classification of SDoH

According to the literature [6-11,22-33], SDoH are associated with COVID-19 transmission and mortality. For example, COVID-19–positive cases or death rates were impacted by SDoH, such as transportation or commuting patterns [29], housing density [22,25], poverty [23,24], health care access [24,27], environmental conditions [6], language barriers [11], occupation [7,23], and residence in rural areas. We argue that determining the correct SDoH variables to measure both health disparities and the spread of diseases is a crucial first step in developing an intelligent surveillance system [34]. While other recent studies have developed health surveillance platforms to facilitate COVID-19 pandemic management and recovery efforts [20,21], our study is unique in its adoption and refinement of the Healthy People 2030's SDoH taxonomy to include and classify SDoH indicators reported previously [6-11,22-33] into the following six domains: (1) SDoH that affect access to resources; (2) SDoH that increase disease exposure, susceptibility, and severity; (3) SDoH that affect adherence to local laws and health policies; (4) SDoH that are community characteristics; (5) SDoH that help increase awareness, knowledge dissemination, and health education; and (6) SDoH specific to neighborhood and built environment that can impact COVID-19–associated comorbidities (Table 1). We used this new SDoH classification as a guideline to collect and analyze the relevant socioeconomic indicators used in UPHO.

**Table 1.** Classification of Social Determinants of Health related to the COVID-19 pandemic.

| Category | SDoH[a] |
|---|---|
| SDoH that affect access to resources | - Access to proper care:<br>  - Distance (miles/hour) to the closest health care facility<br>  - Transportation burden index<br>- Access to healthy food sources:<br>  - Distance to the nearest food market<br>  - Proportion of people without access to a vehicle |
| SDoH that increase disease exposure | - Transportation:<br>  - Proportion of people relying on public transportation<br>  - Proportion of people relying on carpooling<br>  - Proportion of people without access to a vehicle<br>  - SafeGraph mobility data<br>- Age groups:<br>  - Dependents under 18, elderly over 65 years of age<br>  - Proportion of single-parent households<br>  - Households with dependents (children and elderly)<br>  - How the different age groups spend their time<br>- Occupation Type:<br>  - Proportion of health care workers<br>  - Proportion of frontline workers<br>  - Proportion of single vs multiple household earners |
| SDoH that affect adherence to laws and policies | - Occupation Type:<br>  - Proportion of health care workers<br>  - Proportion of frontline workers<br>  - Proportion of single-parent households<br>- Population Density:<br>  - Count of housing units<br>  - Average household size<br>  - Multifamily vs single-family residences |
| SDoH that are community characteristics | - Race, ethnicity, and immigration status<br>  - Proportion of ethnic minorities<br>  - Proportion of racial minorities<br>  - Proportion of first-generation immigrants<br>- Neighborhood quality:<br>  - Social deprivation index<br>  - Blight rating<br>  - Proportion of people under the federal poverty line<br>  - Proportion of unemployment<br>  - Proportion of current or previously incarcerated people<br>- Environmental or safety factors:<br>  - Crime rates<br>  - Distance to parks and community centers<br>  - Distance to police or fire stations<br>  - Proportion of green space coverage<br>  - Air quality index |
| SDoH that enable increasing awareness, knowledge dissemination, and health education | - Digital access and digital inclusion:<br>  - Proportion of people who have access to Wi-Fi or the internet<br>  - Proportion of cellphone or smartphone users<br>- Communication and language barriers:<br>  - Proportion of first-generation immigrants<br>  - Proportion of literate people<br>- Education attainment:<br>  - Proportion of people with a high school diploma<br>  - Proportion of people with a 2-year college diploma<br>  - Proportion of people with a baccalaureate diploma |

| Category | SDoH[a] |
|---|---|
| SDoH specific to neighborhood and built environment that can impact COVID-19 associated co-morbidities | • Parcel or building characteristics<br>• Social deprivation index<br>• Blight rating<br>• Proportion of residential addresses with backyards<br>• Distance to parks and community centers<br>• Proportion of green space coverage<br>• Crime rates<br>• Distance to market/fresh produce<br>• Proportion of smokers |

[a]SDoH: Social Determinants of Health.

## UPHO: A Global Platform

We intend for the UPHO system to impact the health care infrastructure and services at multiple levels; for example, availability of timely, accurate, and complete public health surveillance systems; patient's engagement in self-care or management; competency in patient-centered diagnosis for COVID-19 and other diseases; and effective policy- and decision-making through the analysis of causal relationships among disease prevention, treatment options, and patient outcomes. The process of building UPHO involves (1) collecting and integrating data from several sources, (2) classifying the collected data into drivers and outcomes, (3) incorporating data science techniques for calculating measurable indicators from the raw variables, and (4) determining the extent to which interventions are identified or developed to mitigate drivers that lead to the undesired outcomes. The UPHO architecture is composed of three different layers (Figure 2): data, analytics, and application. The analytics layer extracts the information from the data layer and the applications layer extracts knowledge from the analytics layer. Here we have provided a detailed description of the design of the UPHO platform.

**Figure 2.** Layered architecture of the Urban Population Health Observatory platform from data to application.

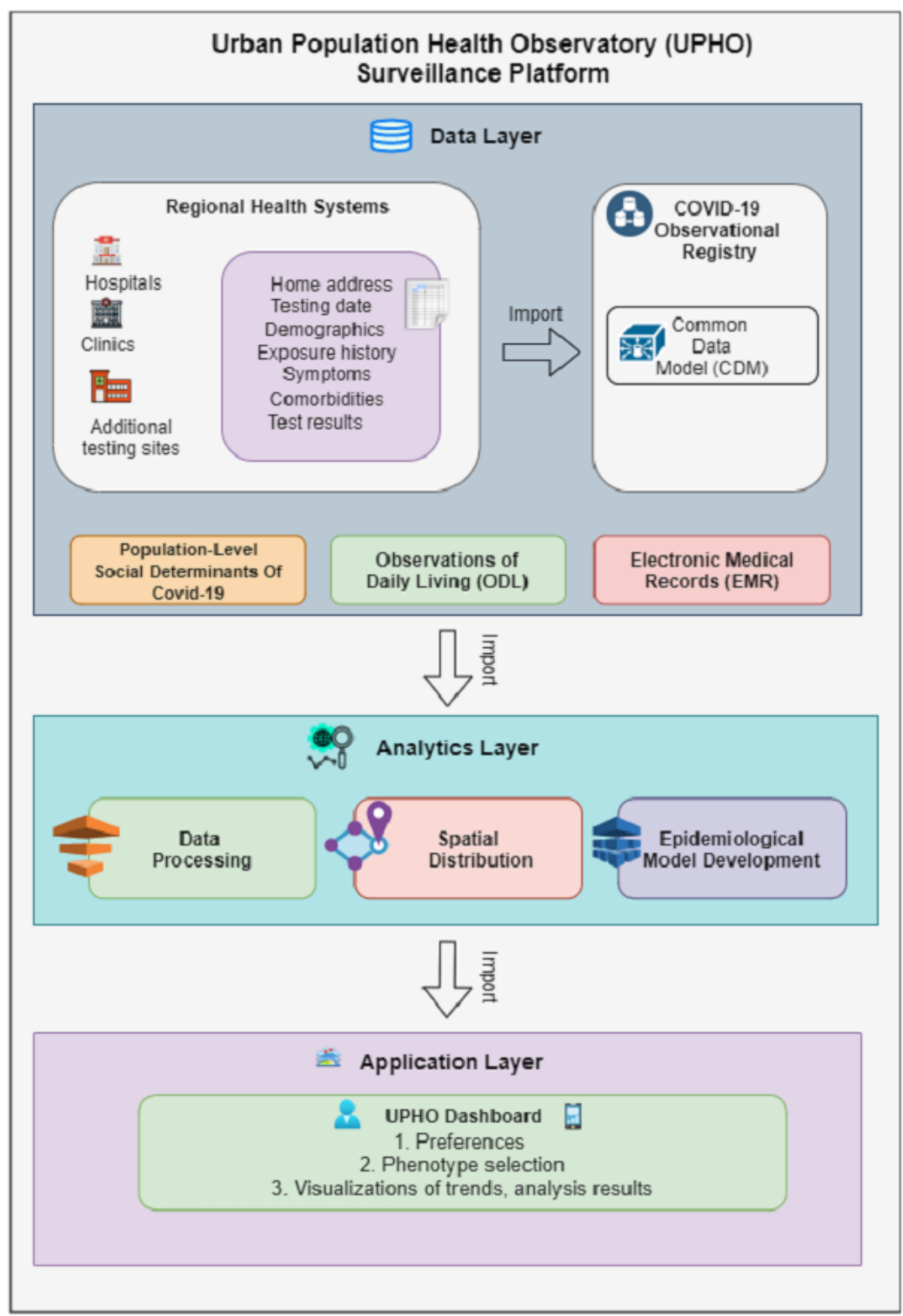

## UPHO Platform Design and Development

### Data Layer

As shown in Figure 2, UPHO integrates data from several sources, including individual-level COVID-19 indicators collected from a regional registry, multidimensional population-level SDoH indicators, SDoH and epidemiological data, clinical data in the patients' electronic medical records, and patient-reported outcomes. These sources are discussed in detail below.

### COVID-19 Observational Registry

The COVID-19 registry systematically collects individual-level COVID-19 indicator variables, including administered test results collected from several testing sites and self-reported outcomes collected from surveys. We performed data transfer between the different sites, including results from both surveys and diagnostic tests stored in the CSV format.

### Population-Level Social Determinants of COVID-19

To obtain SDoH variables, we utilized data from the United States Census Bureau's 2018 American Community Survey [35], US Department of Agriculture, and PolicyMap [36] to obtain the variables in Table 1 at the level of the postal code, census tract, and census block group.

### Individual-Level Observations of Daily Living

Individual-level anonymized data are collected through Internet of Things devices, including wearable devices or mobile global positioning system devices (SafeGraph [37] March 2020-current) that capture adherence to social distancing and shelter in place interventions. We collected social distancing metrics because social distancing and shelter-in-place orders were among the most effective early interventions during the pandemic. For that purpose, we utilized the publicly available SafeGraph [37] movement behavior data set, considering that the phased interventions started from March 30, 2020, through a phased reopening, including how often people visit specifically categorized public locations, the duration of their stay, and where they come from. This publicly available resource is collected anonymously from personal mobile phone usage. We utilized the data set to assess relationships among population movement behavior, transportation patterns, and COVID-19 transmission rates.

### Electronic Medical Records (From Regional Hospitals)

This consists of COVID-19–associated clinical data including infection rates, diagnosis with other chronic conditions or comorbidities (eg, cancer, diabetes, and pregnancy), need for mechanical ventilation, escalation to the intensive care unit, and death.

### Analytics Layer

A core component of UPHO is the analytics layer, which is used for intelligent query-answering to formally propose hypothesis-driven research questions. In the following sections, we explain the different steps that we performed within the analytics layer.

#### *Steps 1 and 2: Data Processing*

Figure 3 provides a schematic representation of the UPHO analytic framework, which comprises a 4-fold process. Step 1 aligns and aggregates individual-level indicators of COVID-19 and population-level social determinants of COVID-19. The consolidated data set constructed in Step 1 is "attribute data" that merged with spatial data in a geographic information systems database in Step 2. These steps pave the way for geospatial intelligence analysis in steps 3 and 4.

**Figure 3.** The Urban Population Health Observatory analytics framework. GIS: geographic information system.

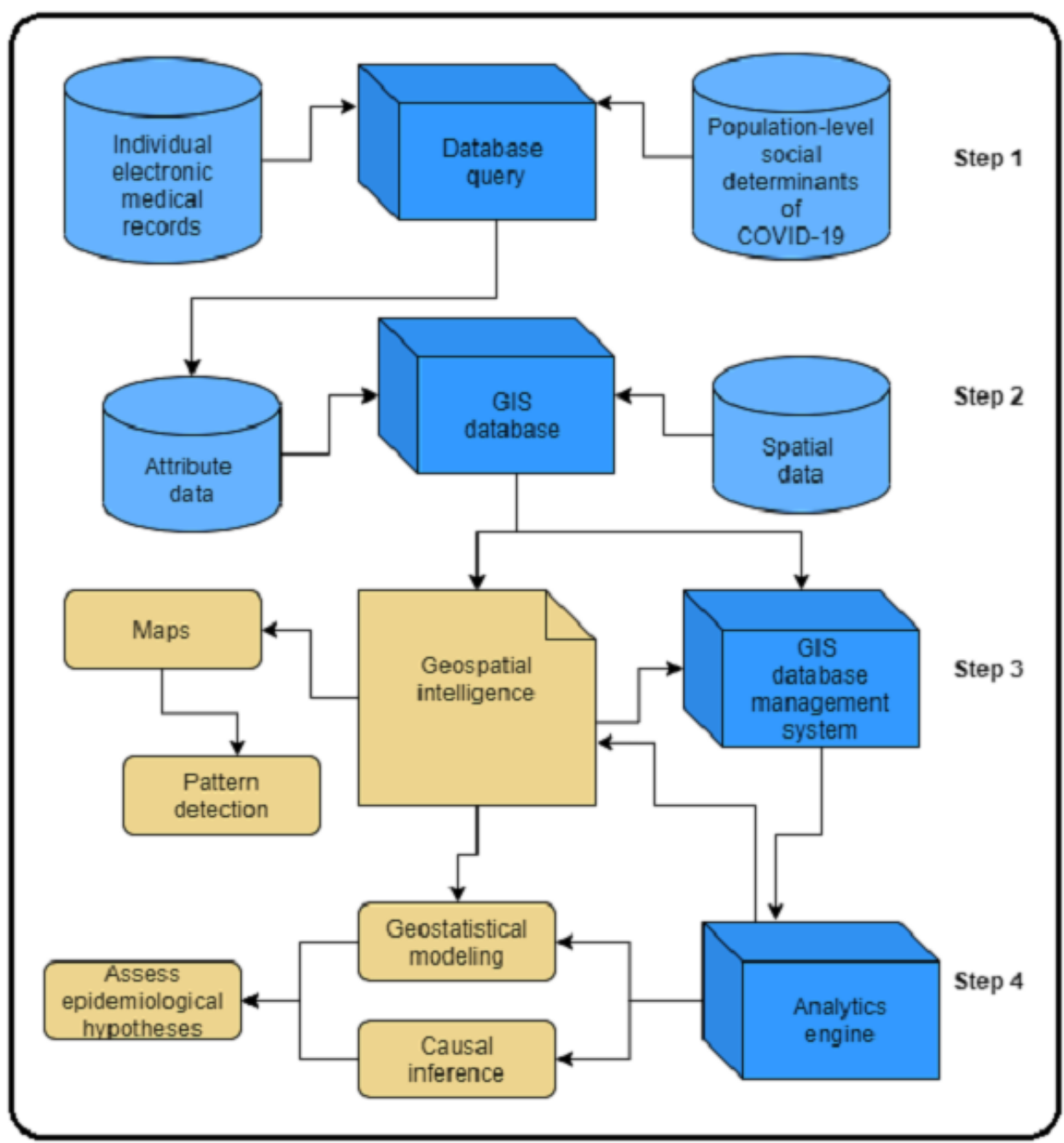

### Step 3: Spatial Distribution

In step 3, we examined the spatial distribution to depict patterns. We conducted geospatial cluster analyses of COVID-19 transmission patterns, which includes neighborhood-level clusters and hotspots to identify high-risk groups. Hotspots are detected by implementing a spatiotemporal pattern mining, which expands the hotspot analysis to 3 dimensions by incorporating time.

### Step 4: Epidemiological Model Development

In this step, we imported data from the geographic information systems' database management system into the analytics engine to perform geostatistical modeling and make causal inferences and to assess various epidemiological hypotheses. We have explained each operation below.

### Step 4.1. Geostatistical Modeling and Causal Inference

**Geostatistical Modeling**

We examined the association among COVID-19 outcomes, SDoH, and policy adherence metrics at various levels of granularity, using global and local geostatistical modeling methods such as ordinary least squares regression and geographically weighted regression (GWR). Among implemented global and local models, we shall also depict which model better explains the association between COVID-19 outcomes and indicators. Global models such as the ordinary least squares regression model (Equation 1) assumes that the processes being modeled are stationary:

$$Y = \beta_0 + \sum_{k=1}^{p} X_k \beta_k + \varepsilon$$

where Y is the COVID-19 outcome variable, $\beta_k$ represents the parameters, $X_k$ represents the observed values of the SDoH and policy adherence metrics variables $k$ (k=1,…,p), and ε represents the random error term.

However, local models such as GWR (Equation 2) generate location-specific results that account for spatial nonstationarity:

$$Y_i = \beta_0(u_i, v_i) + \sum_{k=1}^{p} X_{ik} \beta_k(u_i, v_i) + \epsilon_i$$

where the term $(u_i, v_i)$ represents the coordinates, $\beta_0(u_i, v_i)$ represents the intercept, and $\beta_k(u_i, v_i)$ and $X_{ik}$ are the parameters and observed values of the independent variable $k$ (k=1,…,p), where $i$ ($i = 1, 2, …,n$) represents the spatial location. β values are estimated using spatial weights. $\varepsilon_i$ is the error term for location $i$.

In other words, the strength of GWR compared to global models is the ability to assess varying spatial associations among COVID-19 outcomes, SDoH, and policy adherence metrics.

**Causal Inference**

Causal inference is an important component of the UPHO analytic framework. One of the most challenging tasks in existing population health surveillance systems is encoding causal epidemiological information [38,39]. We utilize causal inference to examine the impacts of SDoH on disease spread and to evaluate the impact of the implemented interventions. Domain knowledge, Spearman rank correlation, and implementation of the Bradford Hill criteria have been used to assess the impacts of SDoH on disease spread, and the Bayesian structural time-series [40] has been used to evaluate the interventions.

**Application Layer**

The application layer within the UPHO architecture (Figure 2) is where we leveraged the inferred knowledge to provide further insights to generate early warnings and inform policymakers. We visualized the results on a dashboard that queries the UPHO through its application programming interface and provides the features explained below.

**Dashboard Prototype**

The analytics layer (Figure 2) performs all analyses offline, including both global and local regression analyses. The results from the analytics layer are stored in a file-based repository that includes different geographic areas at different levels of granularity (eg, postal code, census tract, and census block group), along with their associated SDoH characteristics. It also includes the coefficient estimates between each SDoH and the different COVID-19 outcomes. The dashboard queries the repository of generated analytical results and renders the results on a tabbed view. The different tabs on the tabbed view reflect the 5 different analytics that users might aim to examine (Figure 4). Here we briefly explain the different features provided by this tabbed view and demonstrate some of them through a use case scenario in the following section.

- *F1: causal structure.* The user can use this tab to explore the causal structure with options of selecting different risk factor categories that reflect the social determinants of COVID-19 classification (Table 1).
- *F2: regression analysis.* The user will use this tab to select a level of granularity and explore both a simple association (global regression analysis) and varying spatial coefficient estimate distribution maps (including local regression analysis and GWR).
- *F3: intervention impact.* The user can use this tab to explore the causal impact of the intervention and temporal trends.
- *F4: hotspot analysis results.* The user can use this tab to explore interactive maps of high-risk groups.
- *F5: geospatial disease distribution.* The user can use this tab to select the geographic granularity of interest (eg, postal code, census tract, and block group) and the upstream social determinant risk factors (eg, crime, poverty, and transportation). This will allow the user to explore spatial variability maps.

**Figure 4.** A tabbed view of the 5 main features provided by the dashboard.

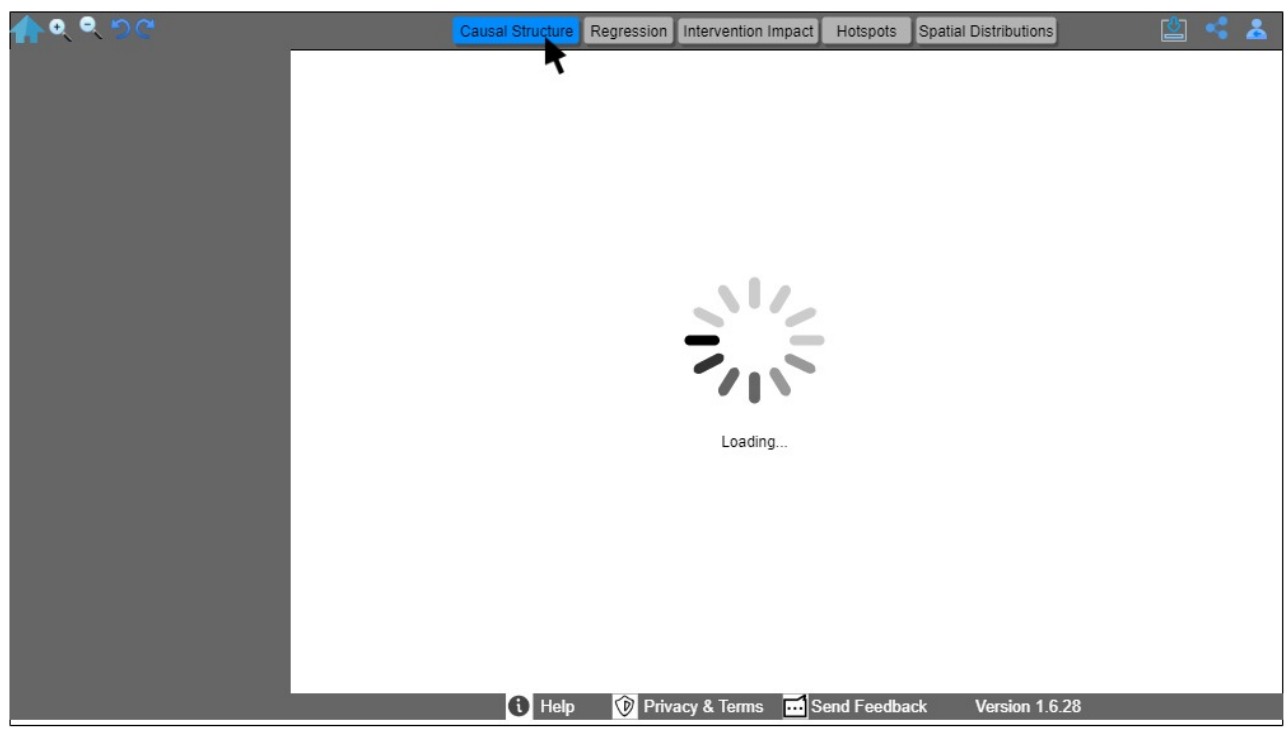

## Use Case Scenarios

In this section, we have provided use case scenarios for each of the 4 different users to demonstrate the features available in the dashboard through the first scenario.

- *Scenario 1:* A public health official wants to launch a task force of physicians and nurses to run mobile testing or vaccination sites that will cater to areas that are socially disadvantaged and have low testing rates to reverse the trend of an infectious disease. Furthermore, she may wish to use neighborhood-level information (eg, the prevalence of positive cases, infection rates, number of testing and vaccination units, average household size, and the prevalence of multigenerational family residences) about temporal trends of COVID-19 to inform the reopening of facilities (eg, schools and restaurants) during or after the pandemic.
- *Scenario 2:* A primary care physician focuses on how SDoH from his patient's neighborhood, such as distance to the nearest health facility, count of housing units, average household size, multigenerational family residence, increased prevalence of frontline workers, and reliance on public transportation, can influence his clinical diagnosis and management plans for patients presenting with COVID-19 symptoms at his clinic.
- *Scenario 3:* A patient uses SDoH characteristics obtained from her neighborhood population health data (eg, distance to the nearest health facility, food market, park or community center, walkability score, and crime rates) to develop self-care and management plans for type 2 diabetes, which renders her at an increased risk of severe COVID-19.
- *Scenario 4:* A caregiver with 2 children uses information from her parents-in-law's neighborhood population health data (eg, the prevalence of positive cases, infection rates, hospitalization rates, and number of testing units) to decide whether to take her children to visit these in-laws for the Thanksgiving holiday.

We used scenario 1 to explore the capabilities of the UPHO surveillance platform and the dashboard features. In this scenario, the health official would be interested in only features F1, F3, and F5, to explore the causal pathways of testing, the geospatial distribution of administered COVID-19 tests and SDoH, and the temporal trend of daily COVID-19 cases in a geographical urban area. We used the UPHO user-centered platform to show how these 3 goals are achieved through the distinctive features of the dashboard.

First, the public health official signs in to the UPHO platform, which will determine her role and establish the appropriate access permissions. On logging in, the health official selects a specific chronic or infectious disease, in this case COVID-19, and the outcome of interest (eg, exploring the number of administered tests, recent cases, hospitalizations, or mortalities). If she selects COVID-19 testing as the outcome of interest, she can explore the causal structure as an analytical aim and the corresponding social determinants of COVID-19 of interest (F1; Figure 5). Through the causal structure, she will see the positive and negative correlations and their relative degrees.

**Figure 5.** Causal structure among COVID-19 testing and Social Determinants of Health within the Urban Population Health Observatory framework.

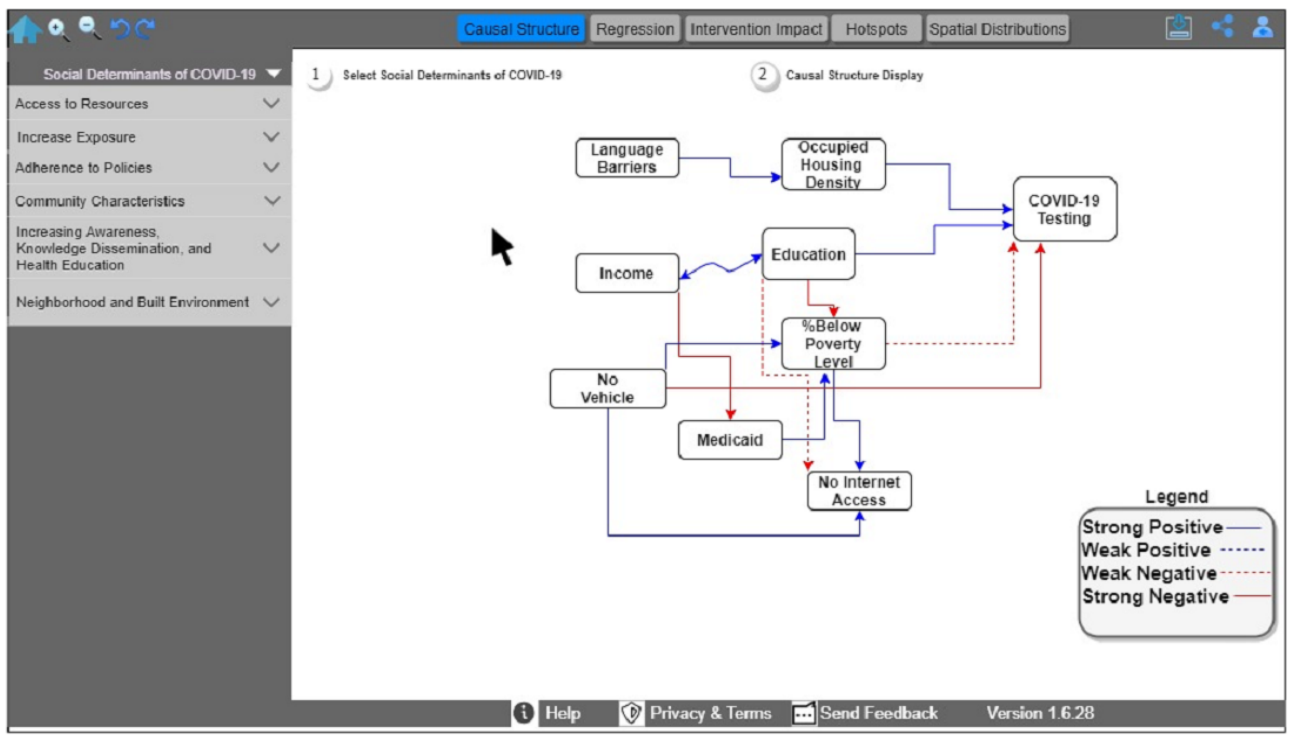

In Figure 5, the causal diagram shows a strong positive relationship among education, occupied housing density, and COVID-19 testing and a strong negative relationship between limited transportation and COVID-19 testing. By selecting new social determinants of COVID-19, the user can explore the causal information encoded in UPHO.

Thereafter, the public health official can explore the administered COVID-19 tests and the SDoH spatial distribution (F5, Figure 6). To that end, she selects the analytics aim of spatial distribution (F5), the specific level of granularity (step 1 in Figure 6), and the social determinant of COVID-19 variable(s) of interest (step 2 in Figure 6). The official can return to the home screen to select the COVID-19 cases as the outcome of interest and then the analytics aim of the intervention impact (F3, Figure 7). She can select "time-series" as a subaim (step 1 in Figure 7), and "daily" as a time measurement (step 2 in Figure 7). She has the option to stratify data by age, sex, or race. The dashboard reflects her selections by displaying an interactive time-series graph, which allows her to hover and explore COVID-19 cases and expand or shrink the graph. According to this example, COVID-19 cases would be declining in the urban area, thus providing evidence in support of the reopening of schools and restaurants in multiple phases while simultaneously considering the necessary safety measures.

**Figure 6.** Spatial distribution of administered COVID-19 tests and Social Determinants of Health.

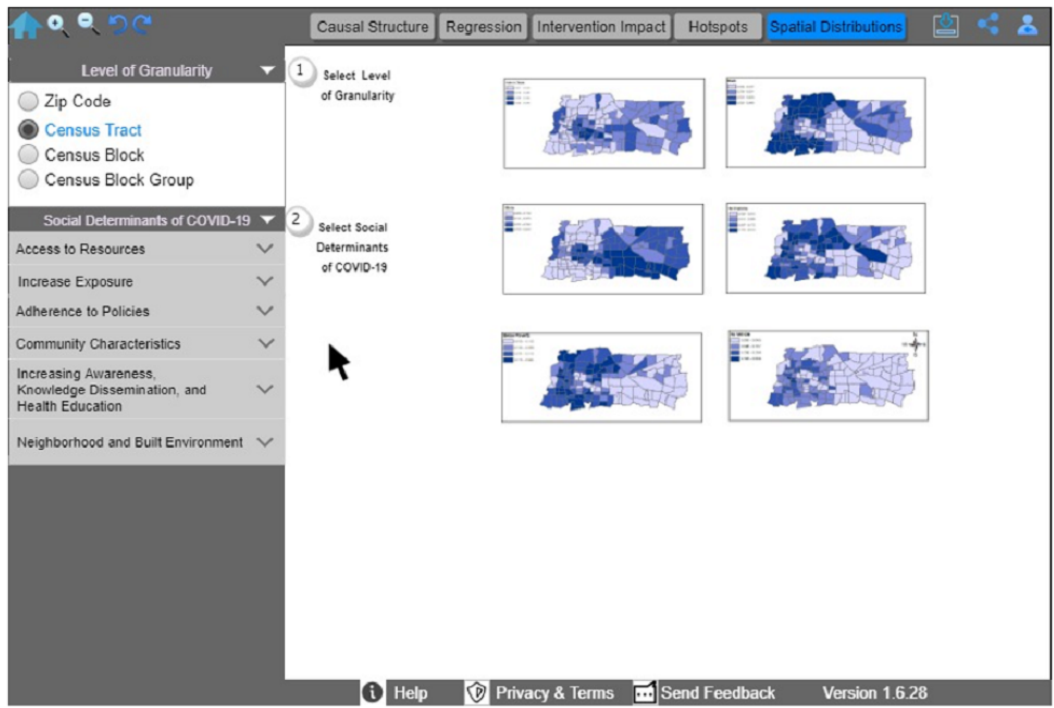

**Figure 7.** An interactive time-series graph that allows users to hover and explore COVID-19 cases, select the time measurement of interest, optionally stratify data on the basis of age, sex, or race, and expand or shrink the graph.

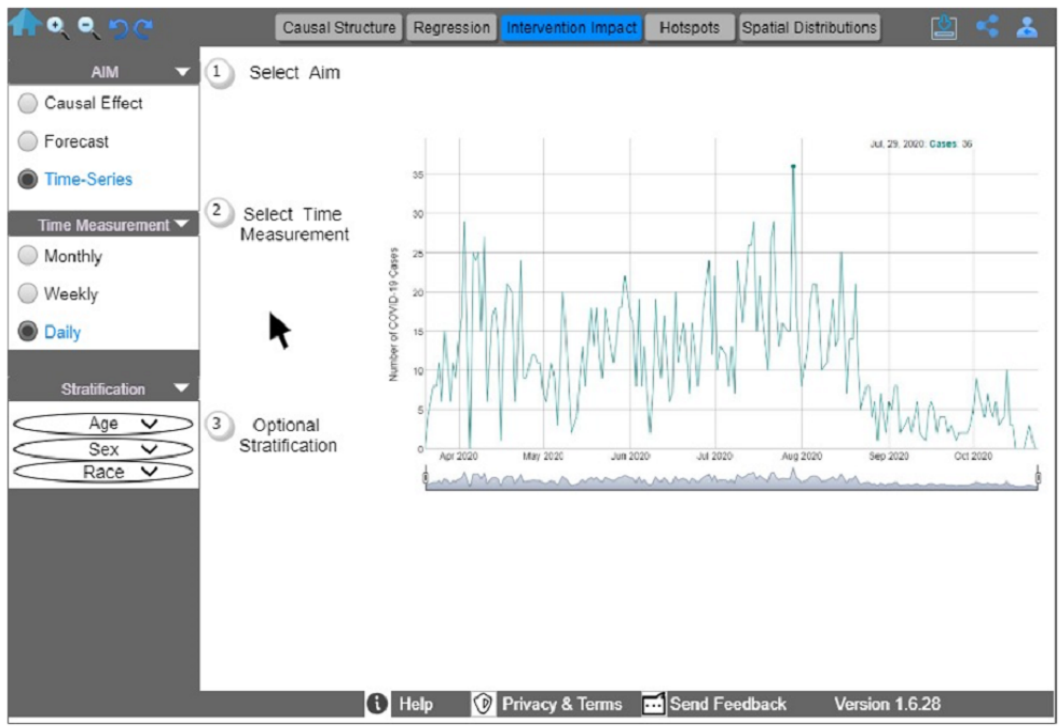

## Discussion

### Principal Findings

Many factors including SDoH increase the burden of COVID-19 on particular groups of the population. In the current study, we proposed the development of UPHO, a web-based urban population health observatory that can help mitigate the disparities of the disease burden of COVID-19 and to facilitate timely responses and better public health planning for equitable distribution of health care resources and services (eg, COVID-19 testing and vaccination). In this study, we redefined the Healthy People 2030's SDoH taxonomy into 6 thematic domains, and these SDoH indicators were integrated with other group-level and individual-level data sources within the UPHO platform. UPHO provides an innovative surveillance tool that systematically incorporates group-level SDoH indicators and population health data to facilitate informed decision-making necessary for preparedness, detection, rapid response, and management after disease outbreaks. The platform provides a reproducible, durable, and scalable model for data-driven, socially informed policymaking for recovery and future-readiness for COVID-19 and other large-scale pandemic events. UPHO enables precision observation and assessment, early detection, and health promotion [41] by disseminating evidence-based, accurate information and facilitating the public's access to timely health information to improve health literacy levels. It can also examine and depict causal pathways between upstream SDoH indicators and COVID-19 outcomes (including other infectious and chronic diseases). By performing local analysis to account for possible local variations, UPHO users can identify location-specific strategies to reduce transmissibility and the burden of diseases and increase the effectiveness of preventive interventions.

Furthermore, depending on the UPHO user's role (eg, public health organizations and primary care physicians) and interests, access can be gained to certain features within the dashboard. For instance, public health officials can visualize graphical representations of analytical results on the dashboard (eg, correlation plots, bar charts, hotspot maps, temporal graphs, and spatial distribution maps). Similarly, physicians may also utilize graphs and charts to summarize and provide context to data from patients' electronic health records, thereby allowing for better clinical decision-making and improved health care efficiency. The dashboard is not limited for use only by scientific investigators, epidemiologists, and health care professionals. Measures of SDoH from the dashboard can be accessible to both the general public and to government officials to identify neighborhood-level risk factors to inform decisions and policymaking. Accordingly, examples of the current and future applications of the UPHO platform include a policymaker who needs to make reliable, accurate, and informed decisions on school reopening and effective policy implementation during a disease outbreak or a clinician who uses information from a patient's postal codes for diagnostic reasoning, monitoring, and disease management. In addition to the epidemiological surveillance of infectious diseases such as COVID-19, the UPHO may also have utility in monitoring and learning about chronic diseases; for example, cancers in the urban population.

The platform implements an intelligent digital health solution that offers the appropriate security and access control mechanisms that are necessary to achieve protections on the privacy of health data.

Although this knowledge-based surveillance platform facilitates intelligent query-answering, there is some potential for improvement that could be considered for future studies. Relationships between upstream risk factors and health outcomes are important, but the addition of contextual knowledge would provide better insights. Consequently, future studies are required to incorporate a semantics layer into the UPHO platform, through which we will define domain ontologies and import some existing ontologies (eg, to study chronic conditions such as obesity [42]) to encode epidemiological knowledge, including concept hierarchies related to health indicators, concepts regarding statistical methods, and causal epidemiological axioms. The ontology will provide contextual knowledge that will help perform semantic inference. Together with the association results, the causal pathways from those inferences can render UPHO an integrated end-to-end analytics platform.

## Conclusions

Social and environmental determinants have a disproportionate impact on minority and disadvantaged populations for COVID-19 infections and related illnesses. Although racial health inequalities have persisted throughout the health care system for years, the COVID-19 pandemic has exacerbated these disparities and made them more visible. Therefore, incorporating SDoH when implementing policies and interventions could facilitate COVID-19 response and management efforts. The UPHO surveillance platform provides a novel approach and novel insights to inform immediate or long-term health policy responses to the COVID-19 pandemic and other future public health crises. In summary, local COVID-19 registries systematically collect individual-level indicators from several regional testing and vaccination sites and align those indicators with population-level indicators, thereby providing decision support, measures for quality care, individual and population-level health reporting systems, and communication tools. The application of the UPHO could help reduce health disparities, thus achieving health equity and improving urban population health.

## Acknowledgments

We would like to thank the Memphis COVID-19 Observational Registry Team led by Prof David Schwartz as well as Drs Fridtjof Thomas, Esra Ozdenerol, and Karen C Johnson for their insights and discussions.

## Conflicts of Interest

None declared.

## Abbreviations

**GWR:** geographically weighted regression
**SDoH:** Social Determinants of Health
**UPHO:** Urban Population Health Observatory

*Edited by T Sanchez; submitted 26.02.21; peer-reviewed by J Ye, G Umeh; comments to author 16.04.21; revised version received 11.05.21; accepted 17.05.21; published 16.06.21*

*Please cite as:*
*Brakefield WS, Ammar N, Olusanya OA, Shaban-Nejad A*
*An Urban Population Health Observatory System to Support COVID-19 Pandemic Preparedness, Response, and Management: Design and Development Study*
*JMIR Public Health Surveill 2021;7(6):e28269*
*URL: https://publichealth.jmir.org/2021/6/e28269*
*doi: 10.2196/28269*
*PMID: 34081605*